\documentclass[conference]{IEEEtran}

\usepackage{cite}
\usepackage{amsmath,amssymb,amsfonts}
\usepackage{algorithmic}
\usepackage{graphicx}
\usepackage{textcomp}
\usepackage[hyphens]{url}
\usepackage{fancyhdr}
\usepackage{algorithm}
\usepackage{subfig} 
\usepackage{booktabs}
\usepackage{pifont}
\usepackage{comment}
\usepackage{fixltx2e}
\usepackage{setspace}
\usepackage{tikz}

\begin{document}

\title{Microarchitecture Design and Benchmarking of Custom SHA-3 Instruction for RISC-V}

\author{
\IEEEauthorblockA{Alperen Bolat, Sakir Sezer, Kieran McLaughlin, Henry Hui}
\IEEEauthorblockA{
{\textit{Queen’s University Belfast}, Belfast, United Kingdom}}
}

\IEEEtitleabstractindextext{%
\begin{abstract}
Integrating cryptographic accelerators into modern CPU architectures presents unique microarchitectural challenges, particularly when extending instruction sets with complex and multistage operations. Hardware-assisted cryptographic instructions, such as Intel’s AES-NI and ARM’s custom instructions for encryption workloads, have demonstrated substantial performance improvements. However, efficient SHA-3 acceleration remains an open problem due to its distinct permutation-based structure and memory access patterns. Existing solutions primarily rely on standalone coprocessors or software optimizations, often avoiding the complexities of direct microarchitectural integration. This study investigates the architectural challenges of embedding a SHA-3 permutation operation as a custom instruction within a general-purpose processor, focusing on pipelined simultaneous execution, storage utilization, and hardware cost. In this paper, we investigated and prototyped a SHA-3 custom instruction for the RISC-V CPU architecture. Using cycle-accurate GEM5 simulations and FPGA prototyping, our results demonstrate performance improvements of up to 8.02× for RISC-V optimized SHA-3 software workloads and up to 46.31× for Keccak-specific software workloads, with only a 15.09\% increase in registers and a 11.51\% increase in LUT utilization. These findings provide critical insights into the feasibility and impact of SHA-3 acceleration at the microarchitectural level, highlighting practical design considerations for future cryptographic instruction set extensions.
\end{abstract}
}

\maketitle
\IEEEpeerreviewmaketitle

\IEEEdisplaynontitleabstractindextext


\section{Introduction and Related Work}

SHA-3 (Secure Hash Algorithm 3), standardized by NIST, is a cryptographic hash function and known for its robust resistance to cryptographic attacks and its adaptability across diverse applications, including authentication, data integrity, and blockchain safety. Unlike previous SHA family members, SHA-3 is built on the Keccak permutation, which employs a sponge construction to absorb input and produce secure hash outputs efficiently. This design underpins SHA-3’s growing adoption in security-critical domains for the future.

To accelerate cryptographic workloads, many modern CPU architectures, such as Intel x86 and ARM, have introduced dedicated instruction set extensions like AES-NI \cite{akdemir2010breakthrough} and ARM Cryptographic Extensions \cite{arm2019cryptographic} that significantly speed up algorithms such as AES and SHA-2. These hardware-assisted instructions have become standard in industry, demonstrating substantial gains in both performance and energy efficiency.

However, despite SHA-3’s increasing adoption in security applications, its integration at the instruction set level remains largely unexplored. Unlike SHA-2, which benefits from well-established instruction-level acceleration, SHA-3 presents unique computational characteristics due to its permutation-based Keccak structure, which consists of extensive bitwise operations, nonlinear transformations, and complex memory access patterns. While numerous studies have explored SHA-3 acceleration via dedicated hardware implementations on ASICs~\cite{gurkaynak2012lessons,nannipieri2021sha2}
or FPGAs~\cite{kobayashi2010prototyping,homsirikamol2011comparing}
. These approaches operate as dedicated accelerators rather than integrated or embedded within a general-purpose processor. The lack of systematic investigation into SHA-3’s microarchitectural integration and integration of dedicated instruction leaves an open research question: how can SHA-3 be efficiently executed as a dedicated instruction within a CPU pipeline, similar to existing industrial solutions like cryptographic instruction extensions proposed for AES and SHA-2?

Prior research has examined alternative SHA-3 acceleration methods, including coprocessors~\cite{azevedo2020sha}, SoC level integrations~\cite{rao2018design}, and modifications on existing standard vector instructions~\cite{li2023maximizing}, as well as optimizations tailored to general-purpose CPU execution pipelines improving compatibility of integration~\cite{sideris2023enhancing} to reduce communication overhead between CPU and accelerator~\cite{bolat2023investigation}. Besides these, some studies have also explored the performance leverage potential of SHA-3 when they closely integrated existing SIMD and Vector instruction in the CPU \cite{cabral2018implementation,rawat2017vector}. However, studies haven't comprehensively analyzed the feasibility, microarchitectural challenges, and trade-offs of embedding dedicated SHA-3 capable custom instruction directly within a CPU’s datapath. Also, key considerations such as pipeline integration, memory efficiency, and hardware resource utilization remain largely unexplored.

In this work, we propose a novel instruction-level acceleration approach for SHA-3 by directly embedding custom SHA-3 (Keccak) capable instruction into a RISC-V processor. We design and implement a SHA-3-specific instruction, analyze its execution within the CPU pipeline, and evaluate its impact on performance and area overhead. Our contributions are as follows:

\begin{itemize}
    \item First microarchitectural study of SHA-3 instruction integration directly into CPU datapath: We analyze the feasibility and challenges of embedding SHA-3 as a dedicated instruction within a general-purpose processor.
    \item Custom SHA-3 instruction design with cycle-accurate CPU analysis and hardware prototype: We design and implement a SHA-3-specific instruction within a RISC-V-based processor on GEM5 and validate on FPGA hardware.
    \item Comprehensive performance evaluation: Using GEM5 simulations and FPGA prototype, we demonstrate significant speedups—up to 8.02× for RISC-V-optimized SHA-3 and 46.31× for Keccak workloads—with minimal hardware overhead (15.09\% increase in flip-flops and 11.51\% increase in LUT utilization).
\end{itemize}

\section{SHA-3 Computational Structure and Keccak-f Permutation Acceleration}

Depending on its operation types, the SHA-3 algorithm consists of two primary components:

\begin{itemize}
    \item \textit {Data Processing}
    \item \textit{Keccak-f Permutation}
\end{itemize}

\textit{Data processing} involves preparing the input message for cryptographic hashing and consists of the following steps:

\begin{itemize} 
    \item \textit{Padding:} A predefined bit sequence is appended to the input message to ensure it is compatible with block size requirements.
    \item \textit{Message Splitting:} The padded message is divided into fixed-sized blocks for sequential processing.
    \item \textit{Absorbing Phase:} Each block is sequentially incorporated into a structured internal state matrix, represented as a \(5 \times 5\) grid of 64-bit words.
\end{itemize} 

These steps primarily involve data movement operations—memory accesses, buffer handling, and state updates—rather than computationally intensive arithmetic or logic operations. Since modern CPU architectures are already optimized for general-purpose data movement and memory operations, accelerating these steps at the instruction level provides limited performance gains.

\textit{Keccak-f permutation}, in contrast, constitutes the computational core of SHA-3. It is responsible for the cryptographic transformation of the state matrix and consists of 24 iterative rounds, each composed of the following five fundamental operations:

\begin{itemize}
    \item \(\theta\) (Theta) – Ensures diffusion by computing parity across column-wise slices of the existing state on calculation.
    \item \(\rho\) (Rho) – Applies cyclic bitwise rotations to introduce non-linearity on state.
    \item \(\pi\) (Pi) – Permutes state bits by rearranging word positions on matrix.
    \item \(\chi\) (Chi) – Applies a non-linear Boolean function based on bitwise operations.
    \item \(\iota\) (Iota) – Injects round-dependent constants to strengthen resistance against linear and differential cryptanalysis.
\end{itemize}

These operations rely heavily on bitwise XORs, ANDs, shifts, and modular rotations, making them computationally expensive when executed sequentially as software on general-purpose CPUs. Unlike the data processing phase, which is primarily constrained by memory bandwidth and register usage, the Keccak-f permutation needs a significant instruction execution overhead due to its reliance on intensive arithmetic and logical calculations.

\subsection{Keccak-f Permutation as Combinational Logic}

A key architectural property of the Keccak-f permutation is that it partially supports implementation as pure combinational logic. 

Since each round of Keccak consists of well-defined bitwise transformations and calculations, dedicated hardware implementations can leverage fully parallelized combinational circuits to compute multiple rounds concurrently, significantly reducing execution latency. This inherent parallelism contrasts sharply with sequential software execution on general-purpose CPUs, where each logical operation turns into multiple instruction cycles, causing high computational overhead. 

While dedicated or co-processor type accelerator designs can exploit this characteristic to achieve ultra-low latency hashing,

modern CPUs lack dedicated hardware primitives for Keccak-f. Instead, they execute the permutation in a highly serialized behavior, requiring multiple instructions for each XOR, shift, and bitwise permutation. This inefficiency motivates the use of custom instruction-level acceleration, wherein dedicated instructions can replace the behavior of combinational logic while remaining within the constraints of a CPU's pipeline.

\begin{figure}[h]
    \centering
    \includegraphics[width=1\linewidth,scale=1]{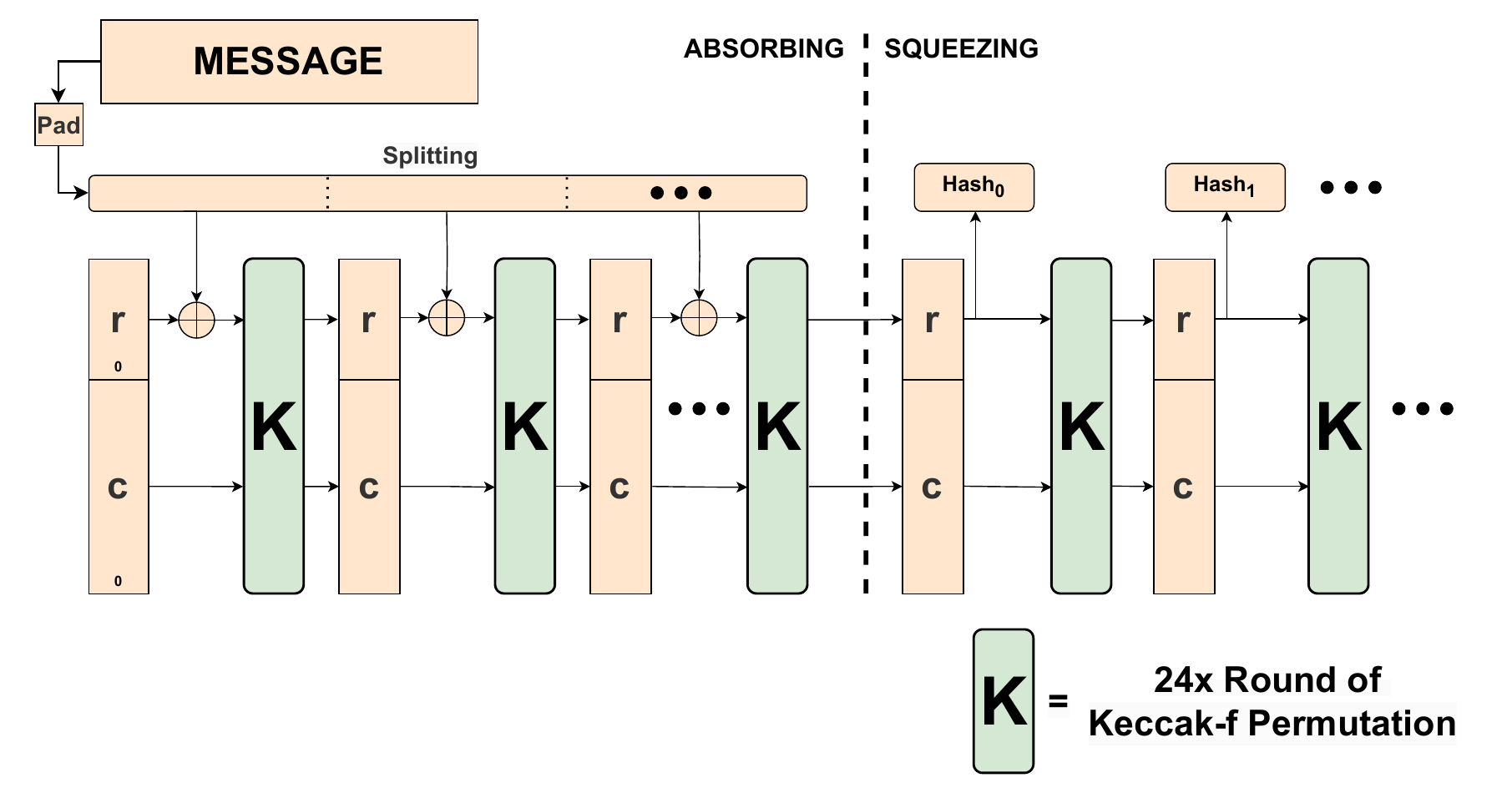}
    \caption{SHA-3 Algorithm Overview}
    \label{fig:sha3_algorithm}
\end{figure}

Figure~\ref{fig:sha3_algorithm} illustrates the SHA-3 algorithm, highlighting the flow and relation between data processing and Keccak-f permutation steps. The parameters, including the rate (\textit{r}) and capacity (\textit{c}), are adjusted based on output length requirements and input data sizes.

\subsection{Motivation for Custom Instruction Acceleration}

In this study, instead of attempting to optimize the entire algorithm, we focus on reducing the Keccak-f permutation's computational complexity, which dominates the total execution time of SHA-3 algorithm. By designing custom instruction that efficiently map the arithmetic and logical transformations involved in \(\theta, \rho, \pi, \chi\), and \(\iota\), we aim to:

\begin{itemize}
    \item \textit{Reduce instruction count:} Minimize the number of general-purpose arithmetic and logic operations required to execute Keccak-f rounds.
    \item \textit{Improve pipeline efficiency:} Enable a tightly coupled execution unit for SHA-3 to process permutation rounds efficiently with fewer cycles and stalls.

    \item \textit{Leverage combinational logic properties:} Utilize custom instructions that align with the parallel nature of Keccak-f, enabling reduced execution latency.
\end{itemize}

This study does not address general-purpose data manipulation tasks, such as padding, message splitting, or absorbing/squeezing phases, as they are inherently constrained by CPU memory architectures rather than computational efficiency. Instead, we focus on investigating accelerating the Keccak-f permutation as a dedicated custom instruction, where the majority of SHA-3's computational effort is concentrated. 

\section{Architectural Design}

The proposed \texttt{shatr} instruction is specifically designed to execute iterations of the Keccak-f permutation in a combinational manner. Each iteration comprises five sequential operations: Theta (\(\theta\)), Rho ( \(\rho\)), Pi (\(\pi\)), Chi (\(\chi\)), and Iota (\(\iota\)). Collectively, these operations implement the cryptographic hashing computations necessary for SHA-3. Given that SHA-3 requires 24 rounds of the Keccak-f permutation per input block, executing the \texttt{shatr} instruction 24 times completes the full permutation process.

Implementations of the Keccak-f permutation iteration can be realized using combinational logic \cite{nannipieri2021sha2,kobayashi2010prototyping,guo2010fair,gurkaynak2012lessons,homsirikamol2011comparing,baldwin2010fpga,ioannou2015high}. According to the standard specification of SHA-3, each input block processed by the Keccak-f permutation consists of 200 bytes of data. Throughout the 24 rounds of the Keccak-f permutation, this data is iteratively updated by performing combinational operations corresponding to each round.

\begin{figure}[h]
    \centering
    \includegraphics[width=\linewidth,scale=1]{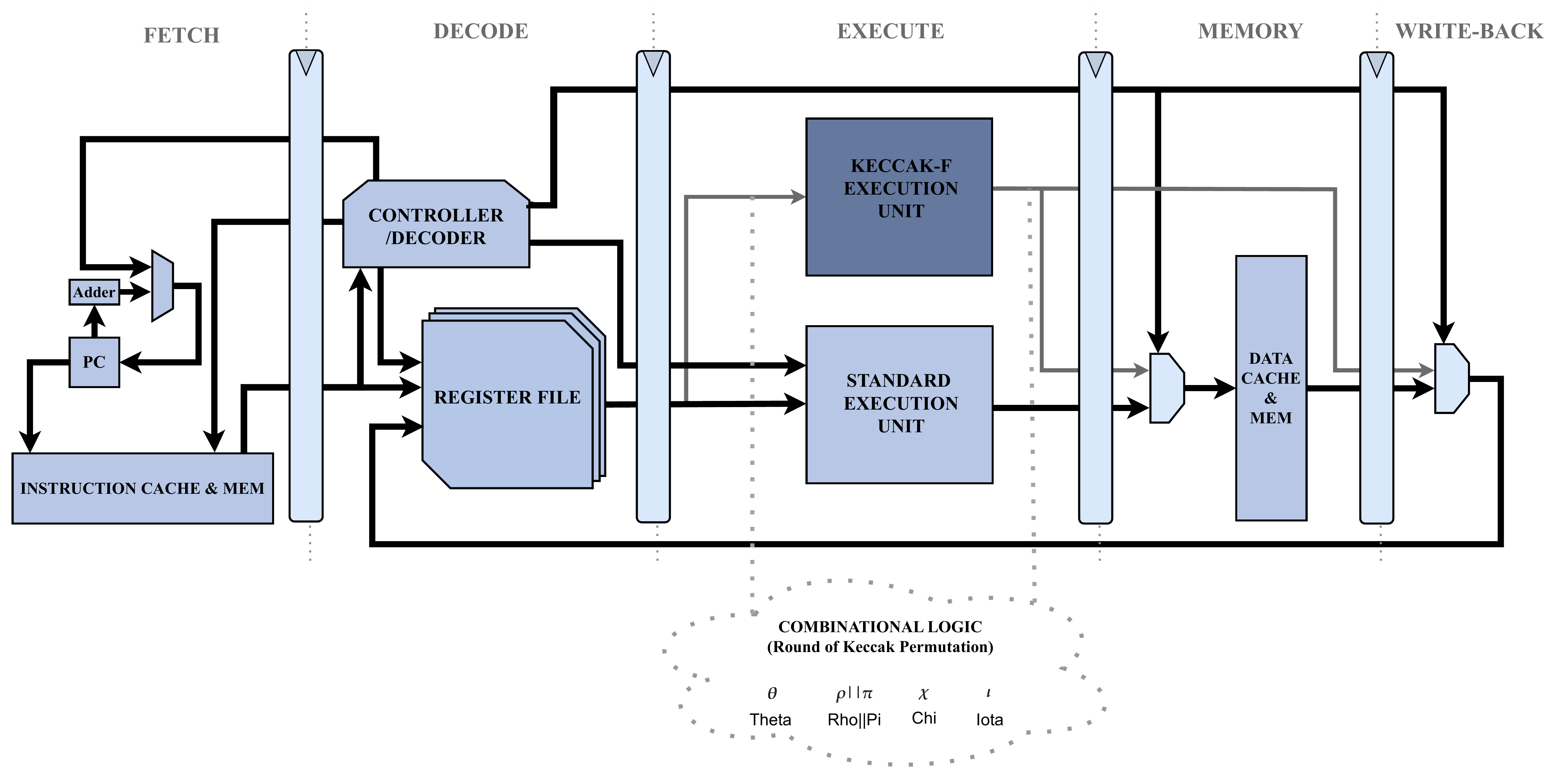}
    \caption{Micro-Architectural Model}
    \label{fig:micro_architecture2}
\end{figure}

To leverage the computational efficiency of the Keccak-f permutation and reduce reliance on standard arithmetic and logical instructions within the CPU datapath, we integrated a dedicated Keccak-f Execution Unit into the CPU pipeline. This execution unit is implemented as a combinational logic-based hardware accelerator specifically optimized for performing Keccak-f permutations. During each execution of the \texttt{shatr} instruction, this unit recursively operates on a 200-byte data chunk. Thus, after executing a single \texttt{shatr} instruction, one round of Keccak-f permutation is completed. Figure~\ref{fig:micro_architecture2} outlines the high-level micro-architecture of the dedicated Keccak-f Execution Unit within a standard CPU pipeline.

By embedding this dedicated Keccak-f Execution Unit into the CPU execution stage, the processor can efficiently perform calculations required by Keccak-f rounds without resorting to multiple standard arithmetic and logical instructions. Consequently, this integration significantly reduces the total number of executed instructions necessary for SHA-3 computations, thereby enhancing overall computational efficiency.

Although standard CPU registers can be utilized to feed data into the Keccak-f Execution Unit, their limited quantity can lead to additional cycles due to frequent data swaps between memory and registers. To address this limitation and improve pipeline parallelism, we implemented dedicated internal registers (flip-flops) totalling 200 bytes within the execution unit itself. These internal buffers function similarly to customized vector registers specifically tailored for buffering intermediate data during iterative Keccak-f permutations. By employing these dedicated internal registers, redundant access to standard CPU registers is minimized or eliminated altogether and with vectorisation, parallel read/write capabilities are achieved in the CPU pipeline.

The proposed \texttt{shatr} instruction does not inherently accelerate data movement nor reduce memory fetch requirements directly. The introduced internal registers are managed exclusively through standard CPU instructions without necessitating additional custom instructions or specialized memory access mechanisms associated with \texttt{shatr} itself. However, by allocating dedicated vector type registers within the CPU pipeline, we effectively mitigate resource constraints typically encountered during Keccak-f computations. This approach indirectly reduces unnecessary register swaps and additional operations required for accessing intermediate data from memory or standard CPU registers.

\begin{figure*}[h]
    \centering
    \includegraphics[width=1\linewidth,scale=1]{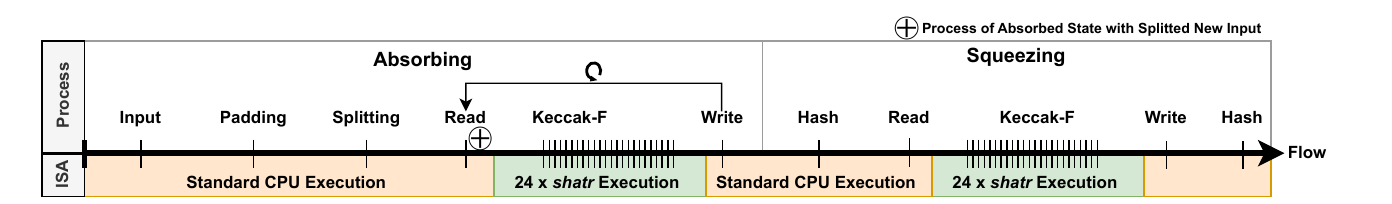}
    \caption{Execution Flow of SHA-3}
    \label{fig:flow_sha3}
\end{figure*}

Figure~\ref{fig:flow_sha3} illustrates the optimized execution flow of SHA-3 enabled by the \texttt{shatr} instruction, emphasizing the reduction in communication overhead achieved through its integration. The figure also addresses the division of execution responsibilities, showing where the Keccak-f Execution Unit, operating with the \texttt{shatr} instruction, performs its specialized combinational tasks and where standard execution occurs during the complete SHA-3 computation process.

\section{Implementation Details}

\subsection{Custom Instruction Integration}

The integration of the \texttt{shatr} instruction required targeted modifications to the RISC-V toolchain, including updates to the assembler, linker, compiler, and simulator, to ensure seamless compatibility with the RISC-V framework. The instruction was assigned to an unused opcode slot, carefully chosen to avoid conflicts with existing instructions while preserving backwards compatibility with the RISC-V Instruction Set Architecture (ISA). This ensures that the addition of \texttt{shatr} does not disrupt standard workloads or require significant changes to the existing software ecosystem.

Once the opcode was assigned, the RISC-V compiler was updated to recognize and generate the \texttt{shatr} instruction during standard compilation. The assembler and linker were modified to support seamless processing of \texttt{shatr} in binary generation, and the simulator was enhanced to accurately model its behavior within the CPU pipeline.

These modifications ensure that \texttt{shatr} integrates seamlessly alongside standard RISC-V instructions while adhering to principles of modularity and backward compatibility. Existing RISC-V applications remain unaffected, while a high-performance path is provided for cryptographic workloads. By embedding this custom instruction into the standard toolchain, the proposed architecture achieves scalability, maintainability, and ease of adoption for both legacy and new software ecosystems. This approach allows developers to leverage the enhanced computational capabilities of \texttt{shatr} without disrupting existing workflows or requiring extensive reengineering efforts.

\subsection{Implementation Details}

To ensure robust evaluation, we used hardware-friendly, C-based implementations of SHA-3 source code aligned with real-world cryptographic workloads and compatible with RISC-V architectures. To evaluate the proposed approach, we selected two software distributions as benchmarks for SHA-3: (i) the official SHA-3 software from the RISC-V organization~\cite{riscvcrypto} and (ii) a standalone SHA-3 implementation published by the Keccak developers~\cite{keccaksoftware}. Both benchmarks were tested using NIST-compliant SHA-3 test vectors~\cite{nistsecurehashing}, covering a range of input sizes, including both short and long messages.

Specifically, we selected two widely recognized distributions: the official implementation provided by the Keccak developers and a RISC-V-optimized variant. These implementations were chosen to establish a reliable baseline and reflect realistic cryptographic processing scenarios. Also, by utilizing RISC-V-based distributions alongside those provided by the Keccak developers, we aimed to ensure that our evaluation avoided reliance on non-RISC-V-optimized software accelerations, which could exaggerate the performance impact of our proposed acceleration. This selection ensures a fair and realistic assessment of the \texttt{shatr} instruction’s impact within RISC-V-specific environments.

The compiling evaluation utilized the latest RISC-V toolchain, configured for the 64-bit instruction set. Customized vector-based registers are exclusively employed to connect the proposed \texttt{shatr} instruction with the combinational logic of the Keccak-f Execution Unit on GEM5. No additional registers, features or optimizations were applied into ISA or CPU, as the objective was to isolate and measure the native performance impact of the custom instruction on these SHA-3 benchmarks.

In both software distributions, we integrated the \texttt{shatr} instruction by mapping SHA-3’s Keccak permutation round operations to \texttt{shatr} through minimal assembly-level modifications of the source code. Only the permutation rounds were altered, leaving the rest of the codebase unchanged. This ensured that any observed performance differences could be attributed solely to the custom instruction, without interference from unrelated optimizations.

Performance results were obtained by comparing \texttt{shatr}-enabled executables against their standard counterparts using a custom cycle-accurate RISC-V CPU model implemented in GEM5. Two configurations were evaluated: one with support for proposed custom \texttt{shatr} instruction and another using a standard RISC-V CPU model without any modification. Memory was configured as 8GB DDR3 with 1600 MT/s and 8 banks.

To further evaluate the hardware utilization introduced by the proposed approach, \texttt{shatr} was configured to work with the RISC-V CVA6 core~\cite{zaruba2019cost} and assessed using an FPGA-based implementation. This analysis provided insights into the impact of the custom instruction on area and resource utilization.

\begin{table*}[h]
\caption{Total Execution Cycles (x10 Million) for RISC-V and Keccak-based Software Distributions of SHA-3 with and without Custom Instruction Across SHA-3 Sizes and Input Lengths}
\centering
\resizebox{0.9\textwidth}{!}{%
\begin{tabular}{|cl|rr|rr|rr|rr|}
\hline
\multicolumn{2}{|c|}{{\textbf{Benchmark}}} &
  \multicolumn{2}{c|}{\textbf{SHA-224}} &
  \multicolumn{2}{c|}{\textbf{SHA-256}} &
  \multicolumn{2}{c|}{\textbf{SHA-384}} &
  \multicolumn{2}{c|}{\textbf{SHA-512}} \\ \cline{3-10} 
\multicolumn{2}{|c|}{} &
  \multicolumn{1}{c|}{\textbf{Short}} &
  \multicolumn{1}{c|}{\textbf{Long}} &
  \multicolumn{1}{c|}{\textbf{Short}} &
  \multicolumn{1}{c|}{\textbf{Long}} &
  \multicolumn{1}{c|}{\textbf{Short}} &
  \multicolumn{1}{c|}{\textbf{Long}} &
  \multicolumn{1}{c|}{\textbf{Short}} &
  \multicolumn{1}{c|}{\textbf{Long}} \\ \hline
\multicolumn{1}{|c|}{{\textbf{\begin{tabular}[c]{@{}c@{}}RISC-V Based \\ Distribution\cite{riscvcrypto}\end{tabular}}}} &
  \textbf{Custom SHA-3 Instruction} &
  \multicolumn{1}{r|}{9.296} &
  436.726 &
  \multicolumn{1}{r|}{8.689} &
  446.975 &
  \multicolumn{1}{r|}{6.331} &
  181.221 &
  \multicolumn{1}{r|}{4.250} &
  147.784 \\ \cline{2-10} 
\multicolumn{1}{|c|}{} &
  \textbf{Without Custom Instruction} &
  \multicolumn{1}{r|}{72.275} &
  3307.817 &
  \multicolumn{1}{r|}{66.993} &
  3434.183 &
  \multicolumn{1}{r|}{51.942} &
  1562.557 &
  \multicolumn{1}{r|}{36.095} &
  1200.756 \\ \hline
\multicolumn{1}{|c|}{{\textbf{\begin{tabular}[c]{@{}c@{}}Keccak-Based\\ Distribution\cite{keccaksoftware}\end{tabular}}}} &
  \textbf{Custom SHA-3 Instruction} &
  \multicolumn{1}{r|}{9.228} &
  440.837 &
  \multicolumn{1}{r|}{8.694} &
  483.317 &
  \multicolumn{1}{r|}{6.384} &
  192.422 &
  \multicolumn{1}{r|}{4.248} &
  157.557 \\ \cline{2-10} 
\multicolumn{1}{|c|}{} &
  \textbf{Without Custom Instruction} &
  \multicolumn{1}{r|}{417.015} &
  19104.412 &
  \multicolumn{1}{r|}{394.303} &
  21313.550 &
  \multicolumn{1}{r|}{297.836} &
  9080.924 &
  \multicolumn{1}{r|}{207.631} &
  7846.874 \\ \hline
\multicolumn{2}{|l|}{\textbf{Total SHA-3 Rounds During Execution of Dataset}} &
  \multicolumn{1}{c|}{\textbf{15,768}} &
  \multicolumn{1}{c|}{\textbf{724,056}} &
  \multicolumn{1}{c|}{\textbf{14,904}} &
  \multicolumn{1}{c|}{\textbf{810,480}} &
  \multicolumn{1}{c|}{\textbf{11,448}} &
  \multicolumn{1}{c|}{\textbf{343,584}} &
  \multicolumn{1}{c|}{\textbf{7,992}} &
  \multicolumn{1}{c|}{\textbf{258,480}} \\ \hline
\end{tabular}%
}
 \label{tab:testenvironment}
\end{table*}

\subsection{Validation of Experimental Environment}

The SHA-3 custom instruction was validated for compatibility and correctness within the RISC-V CPU using the GEM5 simulator. The stability of \texttt{shatr} integration was confirmed through RISC-V unit tests \cite{riscvtests} to ensure ISA compliance. Additionally, benchmarks such as Dhrystone, Linux workloads, and MiBench~\cite{guthaus2001mibench} demonstrated stable performance under computational load. For functional accuracy, hash outputs from the custom instruction were compared against NIST test vectors for all SHA-3 variants, confirming bit-level precision. These results verified robust and efficient operation of the custom instruction without errors or performance degradation.

\section{Experimental Results}

\subsection{Performance Benchmarking}

We benchmarked the custom \texttt{shatr} instruction using two state-of-the-art SHA-3 software distributions: (i) the RISC-V optimized SHA-3 software~\cite{riscvcrypto} and (ii) the Keccak Team’s SHA-3 software implementation~\cite{keccaksoftware}
on GEM5. 
Performance was measured across all SHA-3 output configurations, using both short and long input vectors.

\begin{figure}[h]
    \centering
    \includegraphics[width=0.8\linewidth,scale=1]{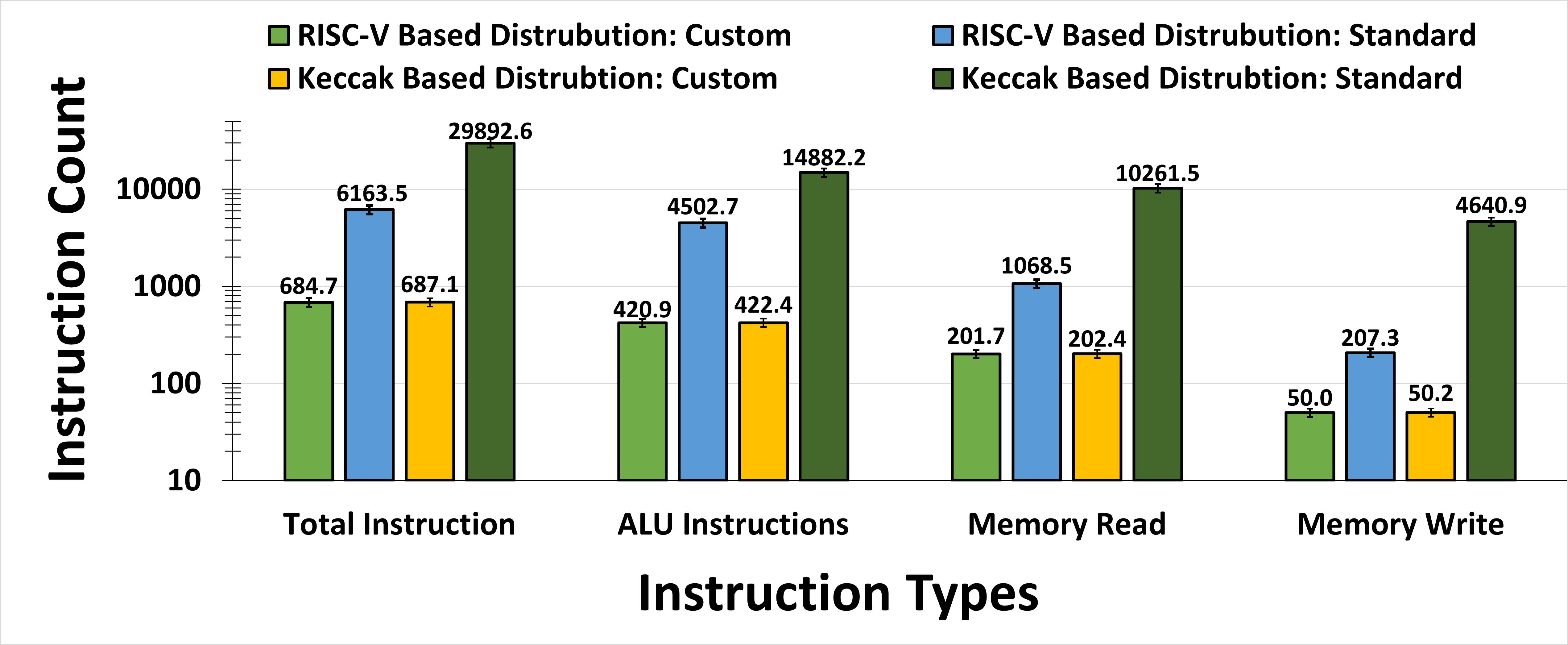} 
    \caption{Comparison of Total Executed Instruction Counts Averages Per SHA-3 Round: Custom Instruction Enabled vs. Standard Execution for Software Benchmarks}
    \label{fig:instruction_count}
\end{figure}

Figure 4 compares the average instruction counts per SHA-3 round, demonstrating that the integration of the custom instruction significantly reduces the number of arithmetic logic unit (ALU) instructions, as well as memory reads and writes. It shows reduced ALU and memory operations per SHA-3 round for both benchmarks. This reduction is particularly pronounced in the Keccak-based software distribution, where the custom \texttt{shatr} instruction substantially minimizes memory request related operations. The implementation of the RISC-V Custom SHA-3 Instruction achieves a notable reduction in ALU-based computational instructions, while maintaining reduced memory access. These results show a significant decrease in memory and bit-wise (arithmetic) instructions per SHA-3 round, following the integration of the \texttt{shatr} instruction.

\begin{figure}[h]
    \centering
    \includegraphics[width=0.8\linewidth,scale=1]{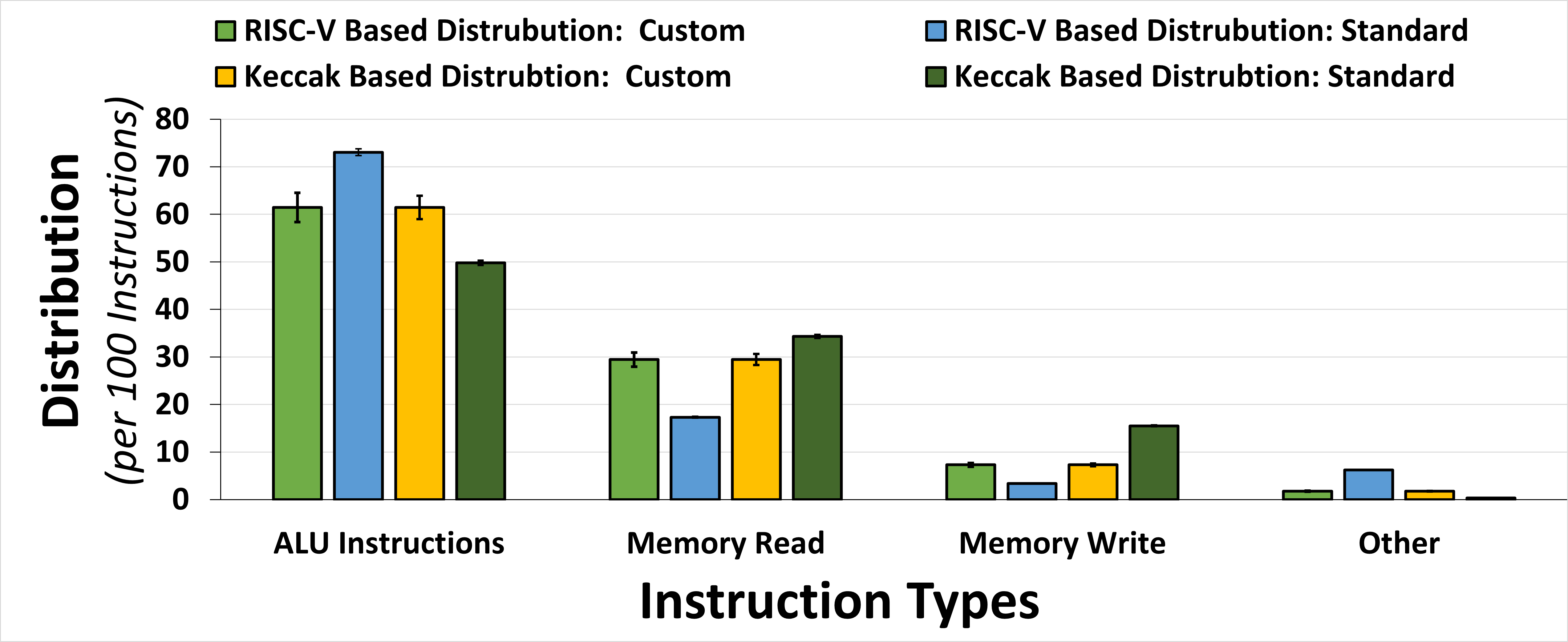} 
    \caption{Instruction Type Breakdown by Percentage: Custom Instruction vs. Standard Execution for Software Benchmarks}
    \label{fig:performance_gain_table}
\end{figure}

Figure 5 presents the distribution of instruction types per 100 instructions, alongside the trend of benchmark changes with the integration of the custom instruction. It reveals that the RISC-V-based benchmark predominantly consists of ALU instructions, while the Keccak-based software primarily utilizes instructions related to memory operations. For both applications, the integration of the custom instruction results in approximately 60\% ALU instructions and a more balanced distribution between the RISC-V-optimized benchmark and the standalone Keccak software. This finding suggests that the RISC-V-optimized software distribution is more compatible with the memory system, whereas the standalone Keccak-based distribution exhibits a higher utilization of memory-related instructions.

\begin{figure}[h]
    \centering
    \includegraphics[width=1\linewidth,scale=1]{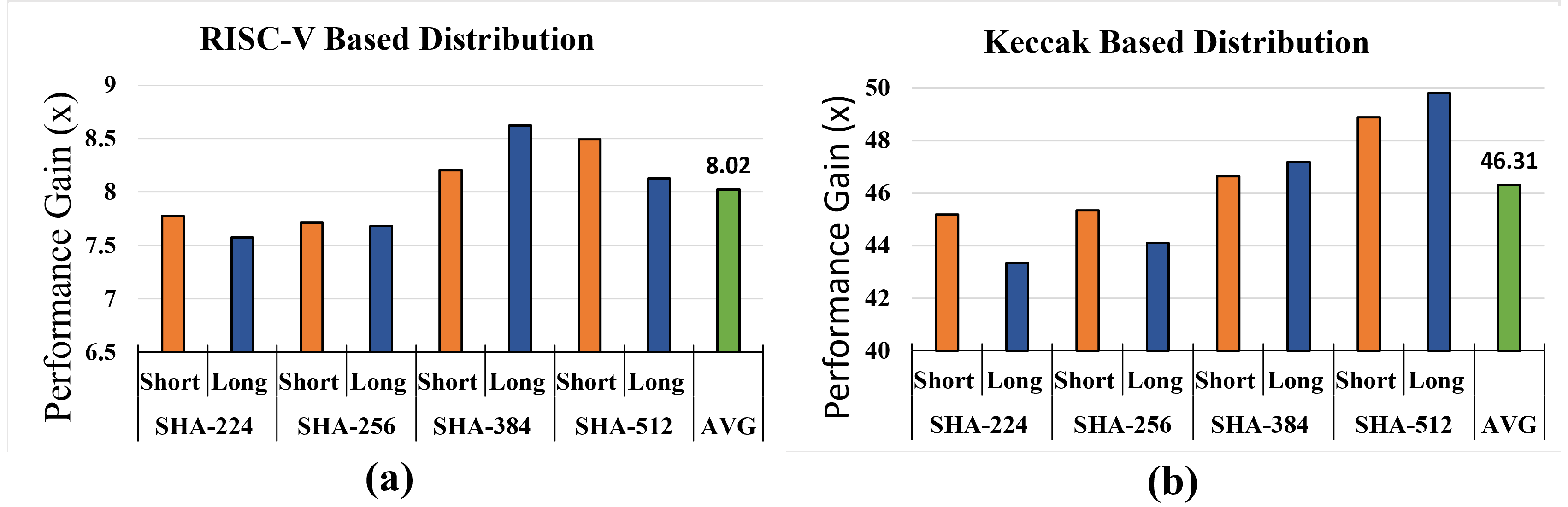} 
    \caption{Normalized Performance Gain in Execution Time: Comparison of Custom Instruction versus Standard (without SHA-3 instruction) for (a) RISC-V Based Software Distribution and (b) Keccak Based Software Distribution}
    \label{fig:performance_gain_table}
\end{figure}

Table I details the total execution cycles for both the RISC-V and Keccak-based software distributions, with and without the custom instruction, across various SHA-3 sizes and input lengths. The data indicates a substantial reduction in execution cycles, particularly for longer input vectors, where the reduction is most pronounced. This demonstrates that the hardware optimization provided by the \texttt{shatr} instruction effectively reduces computational overhead, as reflected by the marked decrease in total cycle counts. Consistent with Figure 4, the Keccak-based SHA-3 distribution is less efficient on RISC-V hardware compared to the RISC-V-optimized distribution, requiring more execution time for SHA-3 processing due to its memory-dominant load characteristics.

The normalized performance gain in execution time is illustrated in Figure 6, which demonstrates a clear improvement when the custom instruction is utilized. Across all SHA-3 sizes, the custom instruction consistently outperforms the standard execution model for both RISC-V and Keccak-based software distributions. While the performance gains are most pronounced for long input vectors with Keccak based SHA-512 and RISC-V based SHA-384, the integration of SHA-3 round instructions led to average improvements of up to 46.31x for Keccak-based software and 8.02x for RISC-V-based software across all workloads. These results confirm that, whether using RISC-V-optimized or standalone SHA-3 software, the incorporation of dedicated SHA-3 round instruction significantly offloads general-purpose SHA-3 execution.

\subsection{Hardware Cost Results}

To validate the efficiency and feasibility of the custom \texttt{shatr} instruction, we implemented and evaluated its hardware realization on an FPGA platform. The design was synthesized and deployed on the Genesys-2 FPGA board, featuring a Xilinx Kintex-7 XC7K325T device. The FPGA implementation assessed area utilization and timing performance. The custom instruction was integrated into a modified RISC-V soft-core processor, implemented in Verilog and synthesized using the Xilinx Vivado toolchain.

\begin{table}[htbp]
  \centering
  \caption{Comparison of FPGA Usage}
  \resizebox{0.8\columnwidth}{!}{
  \begin{tabular}{llll}
    \textbf{Resource} & \textbf{CPU+\texttt{shatr}} & \textbf{CPU Only} 
    & \textbf{Change}(\%) \\ \hline
    Total LUTs & 81,339 & 71,976 & +11.51\% \\ 
    \quad Logic LUTs & \quad79,700 & \quad70,382 & \quad+11.69\% \\ 
    \quad LUTRAMs & \quad1,264 & \quad1,224 & \quad+3.16\% \\ 
    \quad SRLs & \quad375 & \quad370 & \quad+1.33\% \\ 
    FFs & 54,223 & 46,041 & +15.09\% \\ 
    RAMB36 & 50 & 50 & 0.00\% \\ 
    RAMB18 & 2 & 2 & 0.00\% \\ 
    DSP48 Blocks & 27 & 27 & 0.00\% \\ 
    CPU Clock Rate & 50 Mhz & 50 Mhz & 0.00\% \\ 
 \end{tabular}}
  \label{tab:fpga_comparison}
\end{table}

Table \ref{tab:fpga_comparison} summarizes the FPGA resource usage for both the standard RISC-V soft-core and the modified version featuring the \texttt{shatr} instruction. The custom instruction introduces minimal overhead, with only a slight increase in LUTs and Flip-Flops, highlighting its efficient hardware integration without a significant hardware cost increase. Also, timing analysis shows that the custom instruction supports the official 50 MHz clock rate of the CVA6~\cite{CVA6} on the Genesys-2 board, with no added critical path latency.

\section{Conclusion}

In this paper, we proposed a new microarchitecture for a SHA-3 custom instruction for RISC-V that can easily be tailored for any CPU architecture. Further, we presented detailed benchmark results of the proposed microarchitecture based on cycle-accurate GEM-5 simulation. In order to validate our design in terms of additional hardware and speed penalty, we developed an FPGA prototype. The experimental results clearly demonstrates that the proposed \texttt{shatr} instruction is able to achieve substantial performance improvements, with speedups ranging from 8.02x to 46.31x when compared with highly optimized software only implementation. The overall increase of hardware cost for implementing the SHA-3 custom instruction within the RISC-V data path requires additional 15.09\% register and 11.51\% LUTs, while maintaining the original critical path latency of the RISC-V core.

\section*{Acknowledgment}
This work is funded by the UK Government through the New Deal for Northern Ireland. The funding is delivered on behalf of the Northern Ireland Office and the Department for Science, Innovation and Technology by Innovate UK. Additionally, this project is funded as part of the Grant for R\&D by Invest NI (Grant: RD08201502).

\ifCLASSOPTIONcaptionsoff
  \newpage
\fi

\bibliographystyle{IEEEtran}
\bibliography{refs.bib}.

\end{document}